# Quasi-polygonal Conformal Mappings for Designing Whispering Gallery Modes of Multiple Direction Emission

**Haoye Qin, Lemiao Yang**
*School of Instrumentation and Optoelectronic Engineering, Beihang University, Beijing 100191, China*

**Abstract:** We propose a series of conformal mappings for designing directional emission whispering gallery modes. The mappings transform circular cavity into quasi-polygonal transformation cavity. Anisotropic emission is demonstrated in the so-designed cavities.

**Introduction**

Transformation optics is based on the idea that medium of inhomogeneous refractive index will dictate the geometric paths of the ray. Transformation optics is now a technique for designing artificial optical material with targeted functions. Whispering gallery modes provide high-Q resonant modes and low model volume [1], which are understood as closed circular beams supported by total internal reflection (TIR) due to the morphology of the cavity [2,3]. The long lifetime of WGM enables its wide and promising application on low-threshold microlasers [1-4], microcavity sensors [5] with high precision, optical communication and integrate photonic circuits [6]. However, the rotational symmetry of conventional circular WGM will always result in isotropic light emission, which is a major drawback on its functionality [7,8]. In the decades, there are considerable efforts on overcoming this intrinsic problem and obtaining directional light emission of WGMs, for example, employing deformed microcavities [9-11], coupled microcavities [12] and scattering-induced unidirectional emission microcavities [13, 14]. The most typical method of deforming the cavity's shape harms the symmetry to achieve directional light emission, but also undermines the advantageous high-Q modes of WGM.

Recently, Kim et al. [7] first propose the method of designing high-Q directional emission WGMs by transformation optics, the cWGMs. cWGMs are conformal WGMs obtained via transformation optics, which usually employs conformal mapping to deform a unit circle, designs the refractive index in the deformed geometry and still restores the relatively high Q-factor of the WGMs. In the following [6] numerically investigates the two geometry parameters' influence on the Q-factor of the cWGM and proposes the quantitative factors to characterize the capability of direction emission of light. The optimized parameters for the resonant mode make possible both high Q-factor and strong bidirectional emission. [15] continues to resolve the limitation of transformation optics in cWGM designing by employing quasi-conformal mapping. It is able to design arbitrary-shaped WGMs and singularity in the refractive index profile is also removed. In practice, the transformation WGMs can be implemented by drilling subwavelength-scale air holes in a dielectric slab or by setting dielectric posts with high refractive indices [16,17]. Experimental verification at microwave frequencies has demonstrated the effectiveness of designing WGMs via transformation optics [7]. There have been, however, limited researches on multiple direction light emission of WGMs. As is mentioned in [7], this multiple directionality can be realized by applying a certain geometric symmetry via conformal mapping.

In this paper, we propose a series of conformal mappings that transform circular cavity into a rounded quasi-polygonal transformation cavity and demonstrate its multiple directionality in light emission. These mappings enable triple, quadra and multiple direction light emission of WGMs in transformation cavity with high Q-factor retained due to its symmetric quasi-polygonal geometry and gradient refractive index profile.

**Quasi-polygon conformal mappings and anisotropic emission**

In this section we describe the proposed a quasi-polygonal conformal mapping series and their potential applications on designing transformation cavities. A function $w = f(z) = u(x,y) + i*v(x,y)$ in the complex plane is defined as conformal when satisfying the requirements for an analytic function, which is governed by the Cauchy-Riemann equations

$$\frac{\partial u}{\partial x} = \frac{\partial v}{\partial y} \text{ and } \frac{\partial v}{\partial x} = -\frac{\partial u}{\partial y}$$

The proposed quasi-polygonal conformal mapping is given by

$$z(w) = \beta(w + \alpha * w^N)$$

where $\alpha$ is a deformation factor and $\beta$ is a positive value scaling the cavity size. $N$ is an integer responsible for the polygonal side number. This conformal mapping describes the transformation from the circular cavity (unit disk) with coordinates $w = u + iv$ onto the quasi-polygon cavity with coordinates $z = x + iy$.

Optical resonance modes in the cavity are described by the solutions of the following 2-dimensional scalar Helmholtz equation

$$[\nabla^2 + n^2(\mathbf{r})k^2]\psi(\mathbf{r}) = 0$$

where $\mathbf{r}$ is referred to as vector $(x, y)$ and $n(\mathbf{r})$ is the refractive index function. The outgoing-wave boundary condition is expressed as

$$\psi(\mathbf{r}) \sim h(\phi, k)\frac{e^{ikr}}{\sqrt{r}} \text{ for } r \to \infty$$

Therefore, $h(\phi, k)$ is the far-field angular distribution of the emission. Conventionally, in the case of a refractive index homogeneous cavity, the refractive index $n(\mathbf{r})$ is given by

$$n(\mathbf{r}) = \begin{cases} n_0, & \mathbf{r} \in \Omega_1 \\ 1, & \mathbf{r} \in \Omega_2 \end{cases},$$

where $\Omega_1$ and $\Omega_2$ represent the inside and outside domain of the cavity, respectively, and $n_0$ is the constant refractive index of the cavity. Here we set $n_0$ as 1.8. However, the transformation cavity has a refractive index profile instead of a constant refractive index, denoted as

$$n(\mathbf{r}) = \begin{cases} n_0 \left|\frac{dz}{dw}\right|^{-1} = n_0|\beta(1+\alpha w^{N-1})|^{-1}, & \mathbf{r} \in \Omega_1 \\ 1 & , \mathbf{r} \in \Omega_2 \end{cases},$$

In the following, we analyze refractive index distribution, geometric shape and resonances of so-designed transformation cavities by employing commercial software (COMSOL Multiphysics).

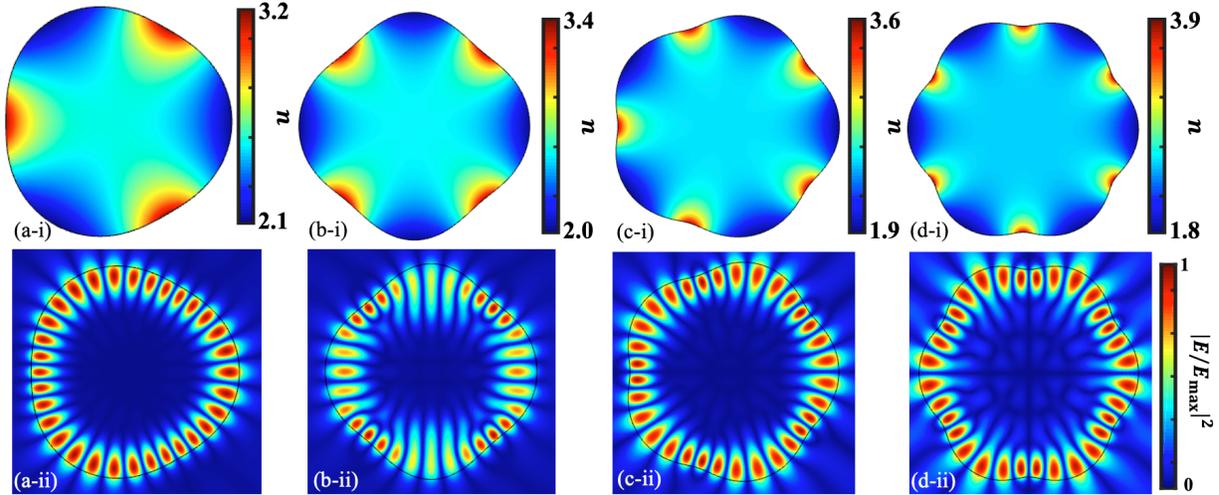

Fig.1 Refractive index profile (i) and near-field (ii) intensity distribution with N=4, 5, 6 and 7.

Fig. 1 shows the refractive index profile of transformation cavity with different number and $(\alpha, \beta)$ fixed at (0.05, 0.714) in the first row and the near-field intensity patterns of each cavity in the second row, respectively. From Fig. 1(a) to Fig. 1(d), the number of the polygonal cavity's side fits the equation of N-1. Like conventional cavity, WGMs in transformation cavities are localized near the boundary. To ensure the condition of total internal reflection is satisfied, $|dz/dw|^{-1} \geq 1$ for determining $\beta$. With a transformation cavity of number N, anisotropic emission in (N-1) directions can be achieved. Here two examples of N=6 and N=7 are demonstrated.

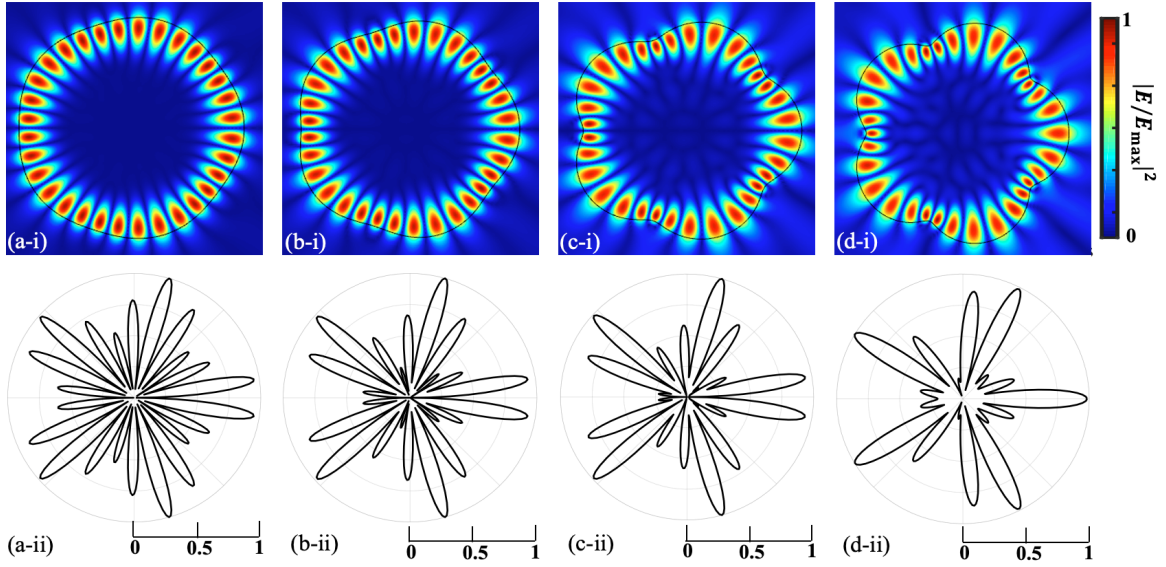

Fig. 2 Near-field (i) and far-field (ii) intensity distribution with $\alpha$=0.02, 0.04, 0.08 and 0.10, $(\beta, N)$=(0.65,6).

Fig. 2 demonstrates the resonance mode near-field distribution and far-field emission pattern with the variation of $\alpha$. The $(\beta, N)$ is fixed at (0.65,6) and $\alpha$ is set to 0.02, 0.04, 0.08 and 0.10. Far-field pattern is more concentrated in the direction with $\alpha$ increasing, and Fig. 2(d-ii) shows a five-direction light emission of the transformation cavity. Q-factors of the four condition are 36065, 7271, 1043 and 499, respectively.

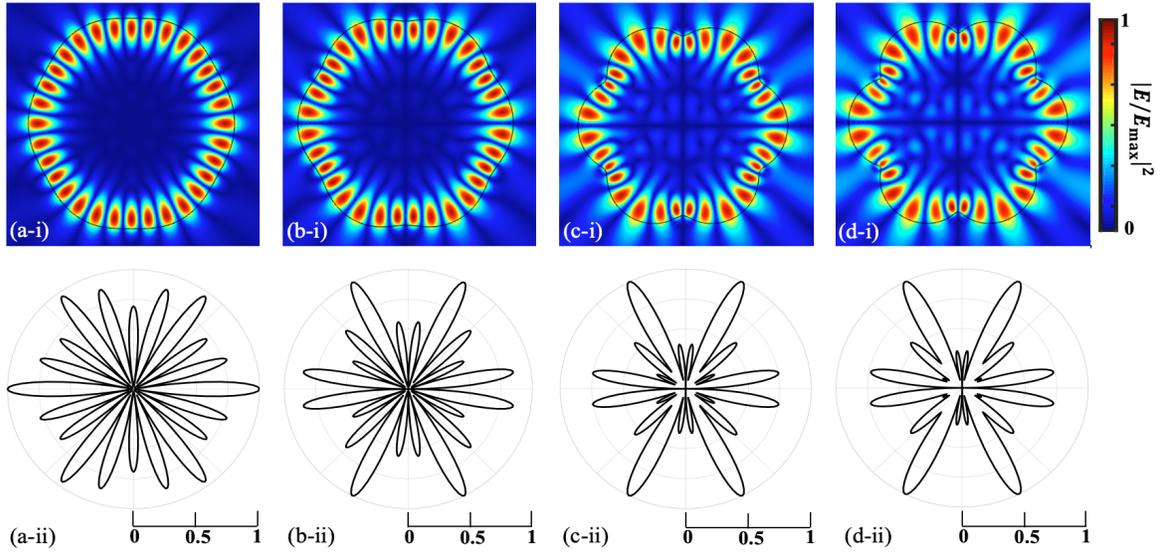

Fig. 3 Near-field (i) and far-field (ii) intensity distribution with $\alpha$=0.02, 0.04, 0.08 and 0.10, $(\beta, N)$=(0.65,7).

Similar to N=6, in Fig. 3 the $(\beta, N)$ is fixed at (0.65,7) and $\alpha$ is set to 0.02, 0.04, 0.08 and 0.10. The resonance modes near-field distribution and their far-field emission pattern are listed as an evolution. From Fig. 3(a-ii) to Fig. 3(d-ii), the side emission is increasingly inhibited and finally in Fig. 3(d-ii) a six-direction light emission condition is achieved. Q-factors of the four condition are 5632, 1367, 318 and 190, respectively.

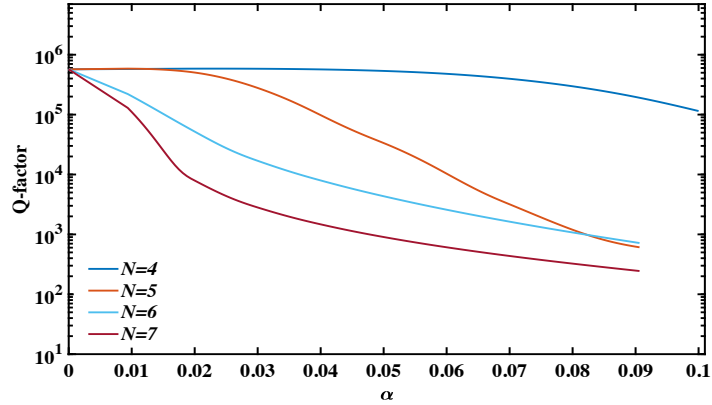

Fig. 4 Q-factor evolution with different $\alpha$ and $N$.

Fig.4 shows the change of Q-factor of WGMs in transformation cavities with different N number and evolution of parameter α. Generally speaking, Q-factor usually experiences a degradation with deformation increasing and at the same α value, a higher N number will result in a lower Q-factor of the transformation cavity.

**Conclusion**

We propose a series of conformal mappings which enable the transformation from a circular homogeneous cavity to a rounded quasi-polygonal transformation cavity. WGMs in the so-designed cavity can achieve triple, quadra and multiple direction light emission. To our best knowledge, this is the first time that multiple direction emission from WGMs is realized via transformation optics.